\documentclass[letter,twocolumn]{jpsj3}
\usepackage{txfonts}
\usepackage{bm}
\usepackage{amsmath}
\usepackage{color}

\def\gsim{\mathop {\vtop {\ialign {##\crcr 
$\hfil \displaystyle {>}\hfil $\crcr \noalign {\kern1pt \nointerlineskip } 
$\,\sim$ \crcr \noalign {\kern1pt}}}}\limits}
\def\lsim{\mathop {\vtop {\ialign {##\crcr 
$\hfil \displaystyle {<}\hfil $\crcr \noalign {\kern1pt \nointerlineskip } 
$\,\,\sim$ \crcr \noalign {\kern1pt}}}}\limits}

\title{Charge Transfer Effect under Odd-Parity Crystalline Electric Field: 
\\
Divergence of Magnetic Toroidal Fluctuation in $\beta$-YbAlB$_4$}

\author{Shinji Watanabe$^1$ and Kazumasa Miyake$^2$}
\inst{$^1$Department of Basic Sciences, Kyushu Institute of Technology, Kitakyushu, Fukuoka 804-8550, Japan \\
$^2$Center for Advanced High Magnetic Field Science, Osaka University, Toyonaka, Osaka 560-0043, Japan
} 

\abst{
A novel property of the quantum critical heavy fermion superconductor $\beta$-YbAlB$_4$ is revealed theoretically.    
By analyzing the crystalline electronic field (CEF) on the basis of the hybridization picture, 
odd parity CEF is shown to exist because of  
sevenfold configuration of B atoms around Yb, which breaks the local inversion symmetry. 
This allows onsite admixture of 4f and 5d wavefunctions with a pure imaginary coefficient, giving rise to the magnetic toroidal (MT) degree of freedom.
By constructing a realistic minimal model for $\beta$-YbAlB$_4$, we show that onsite 4f-5d Coulomb repulsion drives charge transfer between the 4f and 5d states at Yb, which makes  
the MT fluctuation as well as the electric dipole fluctuation diverge simultaneously with the critical Yb-valence fluctuation at the quantum critical point of the valence transition. 
}

\begin{document}
\maketitle

Quantum critical phenomena not following the conventional magnetic criticality~\cite{Moriya,Hertz,Millis} have attracted much attention in condensed-matter physics. 
The heavy-electron metal $\beta$-YbAlB$_4$ with an intermediate valence of the Yb ion~\cite{Okawa2009} exhibits a new type of quantum criticality as the magnetic susceptibility $\chi(T)\sim T^{-0.5}$, the specific-heat coefficient $C(T)/T\sim-\log{T}$, and resistivity $\rho(T)\sim T^{0.5}$ $(T)$ 
for $T\lsim 1$~K ($T\gsim 1$~K)~\cite{Nakatsuji}. 
Furthermore, the $T/B$ scaling where $\chi$ is expressed as a single scaling function of the ratio of 
the temperature $T$ and magnetic field $B$ over four decades was discovered~\cite{Matsumoto}. 
These novel phenomena motivated theoretical studies~\cite{NC2009,WM2010,WM2014,RC2014}. The theory of critical Yb-valence fluctuation (CVF) explains not only the quantum criticality in each physical quantity but also the $T/B$ scaling in a unified way~\cite{WM2010,WM2014}.

Recently, the experimental evidence of the quantum valence criticality has been discovered in 
$\alpha$-YbAl$_{1-x}$Fe$_x$B$_4$ ($x=0.014$)~\cite{Kuga2018}. The sister compound $\alpha$-YbAlB$_4$ shows the Fermi-liquid behavior at low temperatures~\cite{Matsumoto}. However, by substituting Fe to Al 1.4\%, the same unconventional criticality and the $T/B$ scaling as those in $\beta$-YbAlB$_4$ emerges. 
A remarkable point is that at $x=0.014$ a sharp Yb-valence crossover occurs that accompanies a sharp change in the volume~\cite{Kuga2018}.  
This provides experimental verification of the theory of the CVF and also valence crossover arising from the quantum critical point (QCP) of the Yb-valence transition~\cite{WM2010,WM2014}. 

The CVF is the charge transfer (CT) fluctuation between the 4f electron at Yb and the conduction electron. 
In this Letter, 
we clarify a unique nature of the CT effect under the odd parity crystalline electric field (CEF) arising from local configuration of atoms around Yb in $\beta$-YbAlB$_4$. 
By constructing the realistic minimal model, 
we reveal that  
onsite 4f-5d Coulomb repulsion drives the CT between the 4f and 5d states at Yb,
which makes  
the magnetic toroidal (MT) fluctuation as well as the electric dipole (ED) fluctuation diverge simultaneously with the CVF at the QCP of the valence transition. 

\begin{figure}[t]
\includegraphics[width=8cm]{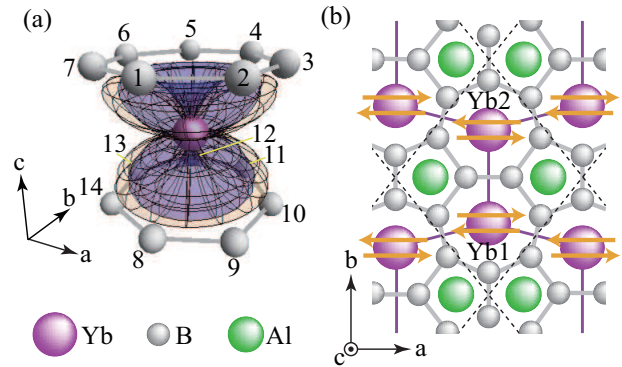}%
\caption{(color online) (a)Yb surrounded by 7 B rings at upper and lower planes. 
The squares of absolute values of spherical parts of the 4f $|J=7/2, J_z=\pm 5/2\rangle$ wavefunction (orange) and 5d $|J=5/2, J_z=\pm 3/2\rangle$ wavefunction (purple) at Yb are also shown. 
(b) Unit cell is enclosed area by dashed lines. Arrows at Yb represent the MT dipole  moments (see text).}
\label{fig:Yb_B}
\end{figure}

Let us start with the CEF in $\beta$-YbAlB$_4$. 
Since in Yb$^{3+}$ the 4f$^{13}$ configuration is realized and in Yb$^{2+}$ the closed shell appears with the 4f$^{14}$ configuration, it is convenient to take the hole picture instead of the electron picture.
The ground state of the CEF for the 4f hole state at Yb has been proposed theoretically to be 
 $|J=\frac{7}{2},J_{z}=\pm\frac{5}{2}\rangle$~\cite{NC2009}, which accounts for the anisoropy of the magnetic susceptibility. 
The realization of the $|J=\frac{7}{2},J_{z}=\pm\frac{5}{2}\rangle$ ground state suggests that 
the CEF in $\beta$-YbAlB$_4$ is to be understood from the hybridization picture rather than the point-charge model. 
This view is compatible with the fact that 
the conical wavefunction of $|J_z=\pm\frac{5}{2}\rangle$ along the $c$ axis spreads almost toward B rings, which acquires the largest 4f-2p hybridization, 
as shown in Fig.~\ref{fig:Yb_B}(a)~\cite{NC2009}.

On the basis of the hybridization picture, let us analyze the CEF at the Yb1 site in the $i$-th unit cell surrounded by seven B atoms at upper and lower planes in Fig.~\ref{fig:Yb_B}(a) [in the unit cell, there are two equivalent Yb atoms labeled by Yb1 and Yb2, respectively, as shown in Fig.~\ref{fig:Yb_B}(b)]. 
The point group symmetry at Yb is approximately sevenfold rotation $C_7$ and hence we take the 2p states at the B site 
as the basis functions: $\varphi_{p_z}$ and $\varphi_{p_{\pm}}\equiv (\varphi_{p_{x}}\pm i\varphi_{p_{y}})/\sqrt{2}$. 
The CEF energy of the $|J_z=\pm\frac{5}{2}\rangle$ state is quantified by the second-order perturbation
$\Delta E(\pm\frac{5}{2},\pm\frac{5}{2})=\langle\pm\frac{5}{2}|\frac{{H_{i1}^{pf}}^{*}H_{i1}^{pf}}{E-H_{0}}|\pm\frac{5}{2}\rangle\equiv\varepsilon_{f}$
with respect to the 4f-2p hybridization 
\begin{eqnarray}
H_{i1}^{pf}=
\sum_{\langle{i1,j}\rangle,m,\sigma,J_z}
\left(
V_{jm\sigma,i1J_z}^{pf}p^{\dagger}_{jm\sigma}f_{i1J_z}+h.c.
\right), 
\end{eqnarray}
where $\langle i1,j\rangle$ represents the nearest-neighbor (N.N.) pairs between B sites and the Yb1 site in the $i$-th unit cell, i.e., $j=1\sim 7$ $(8\sim 14)$ for the upper (lower) plane in Fig.~\ref{fig:Yb_B}(a). 
Here, $m=z, \pm$, $\sigma=\uparrow, \downarrow$, and $J_z=\pm\frac{5}{2}$.

The microscopic origin of the CVF is considered to be the onsite 4f-5d Coulomb repulsion $U_{\rm fd}$ at Yb. Actually, the first-principles calculation shows that the Yb 5d state contributes to the energy band near the Fermi level $E_{\rm F}$~\cite{Suzuki2018}. 
Since the 5d bands are shifted down and forming the wide bandwidth more than 10~eV around $E_F$~\cite{Suzuki2018}, the high-energy 5d-electron state in an isolate Yb ion within the atomic picture i.e. Hund's rule seems to be relevant to the low-energy state in the hole picture in the crystal. As shown in Fig.~\ref{fig:Yb_B}(a), the 5d wavefunction for $|J=5/2, J_z=\pm 3/2\rangle$ spreads most closely to the direction to the B rings, which can acquire the largest 5d-2p hybridization among the $J=5/2$ and $J=3/2$ manifolds. This is compatible with the hybridization picture mentioned above.
Therefore, we proceed our analysis 
based on the hybridization picture 
by considering the 4f $|J_z=\pm\frac{5}{2}\rangle$ state and 5d $|J_z=\pm\frac{3}{2}\rangle$ state at Yb site, which will be denoted as $|\pm\frac{5}{2}\rangle_{4f}$ and $|\pm\frac{3}{2}\rangle_{5d}$, respectively, below.  

Since the local inversion symmetry at the Yb site is broken owing to the sevenfold configuration of B atoms [see Fig.~\ref{fig:Yb_B}(b)], the odd-parity CEF term can arise. This can be explicitly shown by calculating the off-diagonal term by the second-order perturbation
$\Delta E(\pm\frac{3}{2},\pm\frac{5}{2})= {}_{5d}\langle\pm\frac{3}{2}|\frac{{H_{i1}^{pd}}^{*}H_{i1}^{pf}}{E-H_{0}}|\pm\frac{5}{2}\rangle_{4f}$ 
with respect to the 5d-2p hybridization 
\begin{eqnarray}
H_{i1}^{pd}=
\sum_{\langle{i1,j}\rangle,m,\sigma,J_z}
\left(
V_{jm\sigma,iJ_z}^{pd}p^{\dagger}_{jm\sigma}d_{i1J_z}+h.c.
\right), 
\end{eqnarray}
for $J_z=\pm\frac{3}{2}$. 
%
In the hole picture, by 
considering the $4f^0$ and an extra 2p 
hole 
in vacant states as the intermediate state, 
$\Delta E(\pm\frac{3}{2},\pm\frac{5}{2})$ can be calculated as 
\begin{eqnarray}
\Delta E(\pm\frac{3}{2},\pm\frac{5}{2})=
-\sum_{j=1}^{14}\sum_{m\sigma}V_{jm\sigma,i1\pm\frac{3}{2}}^{pd*}
V_{jm\sigma,i1\pm\frac{5}{2}}^{pf}
\frac{
1-n_{jm\sigma}^p
}{\Delta_0+\epsilon_{jm\sigma}^{p}},
\nonumber
\end{eqnarray}
where $\Delta_0 (>0)$ is the excitation energy to the $4f^0$-hole 
state 
(i.e., 4$f^{14}$ electron state),
 and 
$n_{jm\sigma}^{p}$ and  $\epsilon_{jm\sigma}^{p}$ are the filling and energy of 2p 
hole,  
respectively. 
By inputting the Slater-Koster parameters~\cite{Slater,Takegahara} to $V_{jm\sigma,i1\pm\frac{3}{2}}^{pd*}
V_{jm\sigma,i1\pm\frac{5}{2}}^{pf}$ and assuming $n^{p}=n_{jm\sigma}^{p}$ and $\epsilon^{p}=\epsilon_{jm\sigma}^{p}$ for simplicity, we find that 
this off-diagonal term is expressed in the form of pure imaginary: $\Delta E(\pm\frac{3}{2},\pm\frac{5}{2})= iA$ with $A$ being a real number. 
We also have $\Delta E(\pm\frac{3}{2},\pm\frac{3}{2})={}_{5d}\langle\pm\frac{3}{2}|\frac{{H_{i1}^{pd}}^{*}H_{i1}^{pd}}{E-H_{0}}|\pm\frac{3}{2}\rangle_{5d}\equiv\varepsilon_{d}$.

Then, by diagonalizing the $2\times 2$ matrix $\Delta E_{\pm}=\varepsilon_{f}|\pm\frac{5}{2}\rangle_{4f}{}_{4f}\langle\pm\frac{5}{2}|
+\varepsilon_{d}|\pm\frac{3}{2}\rangle_{5d}{}_{5d}\langle\pm\frac{3}{2}|
+iA|\pm\frac{3}{2}\rangle_{5d}{}_{4f}\langle\pm\frac{5}{2}|
-iA|\pm\frac{5}{2}\rangle_{4f}{}_{5d}\langle\pm\frac{3}{2}|
$, 
the CEF ground state is obtained as the Kramers doublet 
\begin{eqnarray}
|\Psi_{\pm}\rangle=\left(u|\pm\frac{5}{2}\rangle_{4f}+Ai|\pm\frac{3}{2}\rangle_{5d}\right)\frac{1}{\sqrt{u^2+A^2}}, 
\label{eq:CEFWF}
\end{eqnarray}
where $u$ is given by $u=(\varepsilon_{f}-\varepsilon_{d}-\sqrt{(\varepsilon_{f}-\varepsilon_{d})^2+4A^2})/2$.  

The admixture of the pure imaginary term in Eq.~(\ref{eq:CEFWF}) is also naturally understood from the local geometry around Yb shown in Fig.~\ref{fig:Yb_B}(b). Since seven B atoms surround each Yb atom symmetrically with respect to the $bc$ plane at which the Yb atom is located, the electric field should work at Yb along the $b$ direction.  If we calculate the ED moment $Q_{i1x}\equiv -ex_{i1}$ and $Q_{i1y}\equiv -ey_{i1}$, whose operators are given by  
$Q_{i1\zeta}=Q_{i1\zeta +}+Q_{i1\zeta -}$ for $\zeta=x, y$ with 
$Q_{i1x+}=
-e
\frac{5}{7}\sqrt{\frac{3}{10}}
d^{\dagger}_{i1+\frac{3}{2}}f_{i1+\frac{5}{2}}+h.c.$, $Q_{i1x-}=
e
\frac{5}{7}\sqrt{\frac{3}{10}}
d^{\dagger}_{i1-\frac{3}{2}}f_{i1-\frac{5}{2}}+h.c.$,
$Q_{1iy+}=-i
e
\frac{5}{7}\sqrt{\frac{3}{10}}
d^{\dagger}_{i1+\frac{3}{2}}f_{i1+\frac{5}{2}}+h.c.$, and 
$Q_{1iy-}=-i
e
\frac{5}{7}\sqrt{\frac{3}{10}}
d^{\dagger}_{i1-\frac{3}{2}}f_{i1-\frac{5}{2}}+h.c.$, 
respectively, by using a general form $|\Psi_{\pm}\rangle=\tilde{u}_{\pm}|\pm\frac{5}{2}\rangle_{4f}+\tilde{v}_{\pm}|\pm\frac{3}{2}\rangle_{5d}$, 
we obtain $\langle\Psi_{\pm}|Q_{i1x}|\Psi_{\pm}\rangle=
e\frac{10}{7}\sqrt{\frac{3}{10}}
{\rm Re}(\tilde{u}_{\pm}^{*}\tilde{v}_{\pm})$ and 
 $\langle\Psi_{\pm}|Q_{i1y}|\Psi_{\pm}\rangle=-
e\frac{10}{7}\sqrt{\frac{3}{10}}
{\rm Re}(i\tilde{u}_{\pm}^{*}\tilde{v}_{\pm})$. 
This implies that when $\tilde{u}_{\pm}$ is a real and $\tilde{v}_{\pm}$ is a pure imaginary, 
we obtain the symmetrically allowed form as 
$\langle\Psi_{\pm}|Q_{i1x}|\Psi_{\pm}\rangle=0$ and $\langle\Psi_{\pm}|Q_{i1y}|\Psi_{\pm}\rangle\ne 0$, 
which is indeed the case of Eq.~(\ref{eq:CEFWF}). 
The ED moment for Eq.~(\ref{eq:CEFWF}) is given by 
$\langle\Psi_{\pm}|Q_{i1y}|\Psi_{\pm}\rangle=
e\frac{10}{7}\sqrt{\frac{3}{10}}
uA/(u^2+A^2)$.

The result of Eq.~(\ref{eq:CEFWF}) indicates that there exists the on-site 4f-5d hybridization, which is usually forbidden in centrosymmetric systems.  The odd-parity CEF term is  
\begin{eqnarray}
H_{i1}^{opCEF}=iA\left(d_{i1+\frac{3}{2}}^{\dagger}f_{i1+\frac{5}{2}}+d_{i1-\frac{3}{2}}^{\dagger}f_{i1-\frac{5}{2}}\right)+h.c.
\label{eq:CEF_odd}
\end{eqnarray} 

Recently, novel multipole degrees of freedom, the MT moment has been defined quantum-mechanically~\cite{HK2018,HYYK2018,WY2017} as  
\begin{eqnarray}
\bm{t}_{l}({\bm r}_i)=\frac{\bm{r}_i}{l+1}\times\left(\frac{2\bm{l}_i}{l+2}+{\boldsymbol \sigma}_{i}\right) 
\label{eq:MT}
\end{eqnarray}
at the site $\bm{r}_i$, 
where $\bm{l}_i$ and ${\boldsymbol \sigma}_i$ are the orbital and spin angular-momentum operators, respectively. 
In the $|\pm\frac{5}{2}\rangle_{4f} \otimes |\pm\frac{3}{2}\rangle_{5d}$ manifold, 
the operators of the MT dipole 
are  derived as $T_{i1\zeta}=T_{i1\zeta +}+T_{i1\zeta -}$ for $\zeta=x, y$ with 
$T_{i1x+}=-i
\mu_{B}
\frac{15}{14}\sqrt{\frac{3}{10}}
f_{i1+\frac{5}{2}}^{\dagger}d_{i1+\frac{3}{2}}+h.c.$, 
$T_{i1x-}=i
\mu_{B}
\frac{15}{14}\sqrt{\frac{3}{10}}
f_{i1-\frac{5}{2}}^{\dagger}d_{i1-\frac{3}{2}}+h.c.$,
$T_{i1y+}=
-
\mu_{B}
\frac{15}{14}\sqrt{\frac{3}{10}}
f_{i1+\frac{5}{2}}^{\dagger}d_{i1+\frac{3}{2}}+h.c.$,   
and  
$T_{i1y-}=
-
\mu_{B}
\frac{15}{14}\sqrt{\frac{3}{10}}
f_{i1-\frac{5}{2}}^{\dagger}d_{i1-\frac{3}{2}}+h.c.$
For the CEF ground state in Eq.~(\ref{eq:CEFWF}) we obtain 
$\langle\Psi_{\pm}|T_{i1x}|\Psi_{\pm}\rangle=\mp 
\mu_{B}
\frac{15}{7}\sqrt{\frac{3}{10}}
uA/(u^2+A^2)$ 
and 
$\langle\Psi_{\pm}|T_{i1y}|\Psi_{\pm}\rangle=0$. 
The MT moments of the Kramers degenerated states are aligned to the $\pm x$~$(\parallel \pm a)$ directions at the Yb1 site as illustrated in Fig.~\ref{fig:Yb_B}(b). 
Since another Yb atom in the unit cell, Yb2, which has opposite sign of $A$ 
in Eq.~(\ref{eq:CEF_odd}), the MT moments are aligned oppositely as shown in Fig.~\ref{fig:Yb_B}(b). At each Yb site, the net MT moment is zero due to the Kramers degeneracy. 

On the basis of these analyses, let us construct the Hamiltonian for the periodic crystal 
of $\beta$-YbAlB$_4$, which consists of the 4f $J_z=\pm 5/2$ and 5d $J_z=\pm 3/2$ states at Yb and $2p_z$ and $2p_{\pm}$ states at B:  
\begin{eqnarray}
H=\sum_{i\alpha}\left[H_{i\alpha}^{f}+H_{i\alpha}^{pf}+H_{i\alpha}^{pd}
+H^{U_{fd}}_{i\alpha}
\right]
+H^{d}
+H^{p},
\label{eq:EPAM}
\end{eqnarray}
where $\alpha=1,2$ specifies the Yb1 and Yb2 sites, respectively. 
Here, the $4f$ part is given by 
%
$
H_{i\alpha}^{f}=\varepsilon_{\rm f}\sum_{J_z=\pm5/2}n^{f}_{i\alpha J_z}
+Un_{i\alpha +5/2}^{f}n_{i\alpha -5/2}^{f},
$ 
%
where $U$ is the onsite Coulomb repulsion. 
The $5d$ part is given by 
%
$
H^{d}=\varepsilon_{\rm d}\sum_{i}\sum_{\alpha=1,2}\sum_{J_z=\pm 3/2}n^{d}_{i\alpha J_z}
+\sum_{\langle{i\alpha,i'\alpha'}\rangle}\sum_{J_z,J_z'=\pm 3/2}t_{i\alpha J_z,i'\alpha'J_{z}'}^{dd}d^{\dagger}_{i\alpha J_z}d_{i'\alpha'J_{z}'},
$ 
%
where $\langle{i\alpha,i'\alpha'}\rangle$ takes the N.N. Yb pairs. 
The 4f-5d Coulomb repulsion at Yb is given by 
\begin{eqnarray}
H_{i\alpha}^{U_{fd}}=U_{fd}\sum_{J_z=\pm5/2}\sum_{J_z'=\pm 3/2}n_{i\alpha J_z}^{f}n_{i\alpha J_z'}^{d}.
\end{eqnarray}
The transfer for 2p states is given by 
%
$
H^{p}=\sum_{\langle{j,j'}\rangle\sigma}\sum_{m,m'=z,\pm}t_{jm,j'm'}^{pp}p^{\dagger}_{jm\sigma}p_{j'm'\sigma},
$
%
where $\langle{j,j'}\rangle$ takes the N.N. B pairs in the $ab$ plane and in the $c$ direction [see Fig.~\ref{fig:Yb_B}(a)]. 

Although $H_{i\alpha}^{opCEF}$ is not included in Eq.~(\ref{eq:EPAM}) explicitly, the effect is expected to inhere via the 2p-4f and 2p-5d hybridizations. 
To clarify the effect of the CT under the odd-parity CEF, we apply the slave-boson mean-field theory~\cite{OM2000} to Eq.~(\ref{eq:EPAM}). 
To describe the state for $U=\infty$ responsible for heavy electrons, we consider $f^{\dagger}_{i\alpha J_z}b_{i\alpha}$ instead of $f^{\dagger}_{i\alpha J_z}$ in Eq.~(\ref{eq:EPAM}) by introducing the slave-boson operator $b_{i\alpha}$ to describe the $f^0$-hole 
state and require the constraint $\sum_{i\alpha}\lambda_{i\alpha}(\sum_{J_z=\pm\frac{5}{2}}n^{f}_{i\alpha J_z}+b_{i\alpha}^{\dagger}b_{i\alpha}-1)$ with $\lambda_{i\alpha}$ being the Lagrange multiplier. 
For $H_{i\alpha}^{U_{fd}}$ in Eq.~(\ref{eq:EPAM}), we employ the mean-field decoupling as $U_{fd}n^{f}_{i\alpha J_{z}}n^{d}_{i\alpha J'_{z}}\simeq U_{fd}\bar{n}^{f}_{\alpha}n^{d}_{i\alpha J'_{z}}+R_{\alpha}n^{f}_{i\alpha J_z}-R_{\alpha}\bar{n}^{f}_{\alpha}$, where $R_{\alpha}\equiv U_{fd}\bar{n}^{d}_{\alpha}$ and  $\bar{n}_{\alpha}^{\eta}\equiv\frac{1}{N}\sum_{iJ_z}\langle n^{\eta}_{i\alpha J_z}\rangle$ ($\eta$=f, d) with $N$ being the number of unit cells. 
By approximating mean fields as uniform ones, i.e., $\bar{b}_{\alpha}=\langle b_{i\alpha}\rangle$ and $\bar{\lambda}_{\alpha}=\lambda_{i\alpha}$, the set of mean-field equations is obtained by $\partial\langle H\rangle/\partial\bar{\lambda}_{\alpha}=0$, $\partial\langle H\rangle/\partial \bar{b}_{\alpha}=0$, and $\partial\langle H\rangle/\partial R_{\alpha}=0$: 
$\frac{1}{N}\sum_{{\bm k}J_z}\langle f_{{\bm k}\alpha J_z}^{\dagger}f_{{\bm k}\alpha J_z}\rangle+\bar{b}_{\alpha}^2=1$, 
$\frac{1}{2N}\sum_{{\bm k}}\left[\sum_{J_{z}\xi m\sigma}V^{pf*}_{{\bm k},\xi m\sigma, \alpha{J_z}}\langle f^{\dagger}_{{\bm k}\alpha{J_z}}p_{{\bm k}\xi m\sigma}\rangle
+h.c.\right]+\bar{\lambda}_{\alpha}\bar{b}_{\alpha}=0$, 
and 
$\bar{n}^{f}_{\alpha}=\frac{1}{N}\sum_{{\bm k}J_z}\langle f^{\dagger}_{{\bm k}\alpha{J_z}}f_{{\bm k}\alpha{J_z}}\rangle$. 
Here, $\xi$ specifies the N.N. B sites for the Yb$\alpha$ site [see Fig.~\ref{fig:Yb_B}(a)]. 
We solve these equations together with the equation for the filling $\bar{n}\equiv\sum_{\alpha=1,2}(\bar{n}^{f}_{\alpha}+\bar{n}^{d}_{\alpha})/4+\sum_{j=1}^{8}\bar{n}^{p}_{j}/16$  
with $\bar{n}^{p}_{j}\equiv\frac{1}{3N}\sum_{{\bm k}m\sigma}\langle p^{\dagger}_{{\bm k}jm\sigma}p_{{\bm k}jm\sigma}\rangle$ self-consistently. 

As for the relation among the Slater-Koster parameters, 
following the argument of the linear combination of atomic orbitals~\cite{Andersen1977}, we set 
$(pp\pi)=-(pp\sigma)/2$, $(pd\pi)=-(pd\sigma)/\sqrt{3}$, $(pf\pi)=-(pf\sigma)/\sqrt{3}$, $(dd\pi)=-2(dd\sigma)/3$, and $(dd\delta)=(dd\sigma)/6$.  
In the hole picture, we take the energy unit as $(pp\sigma)=-1.0$ and set   
$(pd\sigma)=0.6$, $(pf\sigma)=-0.3$, $(dd\sigma)=0.4$ as typical values. 
We found that the calculated band structure for $\varepsilon_{d}=-1$ and $\varepsilon_f\approx 
-2.3
$ at $\bar{n}=23/32$ well reproduces the recent photoemission data near $E_{F}$~\cite{Cedric} and 
then here we employ these parameters. 
We performed numerical calculations in the ground state in the $N=8^3, 16^3$ and $32^3$ systems and the results in $N=32^3$ will be shown below. We confirmed that Eq.~(\ref{eq:CEFWF}) with these parameters reproduces the tendency of the anisotropy in the magnetic susceptibility observed in $\beta$-YbAlB$_4$~\cite{Matsumoto,Matsumoto2011}.  

\begin{figure}[t]
\includegraphics[width=8cm]{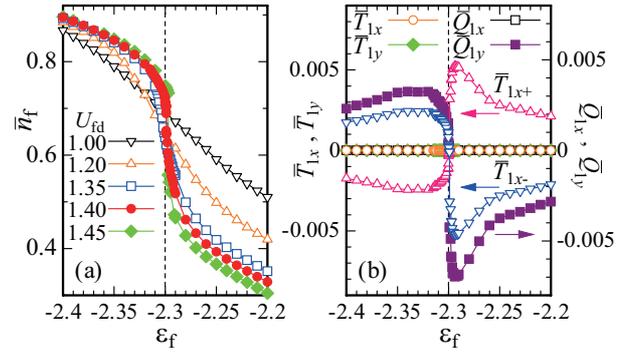}%
\caption{(color online) 
(a) The 4f-hole number $\bar{n}_f$ at Yb vs. $\varepsilon_f$ for $U_{fd}=
1.00
$, 
$
1.20
$, 
$
1.35
$, 
$
1.40
$, and 
$
1.45
$. (b) The $\varepsilon_f$ dependences of the MT moment and ED moment at the Yb1 site for 
$
U_{fd}=1.40
$. We set $e=1$ $(\mu_{B}=1)$ for 
the plot of 
the ED (MT) moment. 
In (a) and (b), the dashed line indicates $\varepsilon_{f}=\varepsilon_{f}^{QCP}$. 
}
\label{fig:nfnd}
\end{figure}

Figure~\ref{fig:nfnd}(a) shows the $\varepsilon_{f}$ dependence of the 4f-hole number per Yb $\bar{n}^{f}(=\bar{n}^{f}_1=\bar{n}^{f}_2)$. 
As $U_{fd}$ increases, $\bar{n}_f$ changes steeply as a function of $\varepsilon_{f}$ and for $U_{fd}=
1.40
$ 
the slope $-\partial\bar{n}^f/\partial\varepsilon_{f}$ diverges at $\varepsilon_{f}=
-2.3001
$. 
Since the valence susceptibility is defined as $\chi_v\equiv -\partial\bar{n}^f/\partial\varepsilon_{f}$, this result indicates that 
the CVF diverges $\chi_v=\infty$ at 
the QCP of the valence transition $(\varepsilon_f^{QCP},U_{fd}^{QCP})\approx
(-2.3001, 1.40)
$. 
For $U_{fd}>U_{fc}^{QCP}$, a jump in $\bar{n}_f$ appears, indicating the first-order valence transition. 

Our realistic minimal model Eq.~(\ref{eq:EPAM}) shows that the QCP with an intermediate valence $\bar{n}^{f}=
0.71
$ as the measurements in $\beta$-YbAlB$_4$~\cite{Okawa2009} and $\alpha$-YbAl$_{1-x}$Fe$_x$B$_4$ (x=0.014)~\cite{Kuga2018} is realized by $U_{fd}^{QCP}=
1.40
$. 
This value is evaluated to be $U_{fd}^{QCP}\approx 
6.6
$~eV, 
if we employ the typical value $(pp\sigma)\approx 4.7$~eV estimated from the band-structure calculation in B~\cite{Papa}.  Since $U_{fd}$ is the onsite interaction, this value seems reasonable. We propose that $U_{fd}$ can be directly examined by recently-developed partial fluorescence yield measurement of the Yb $L_3$ edge~\cite{Tonai2017} in $\beta$-YbAlB$_4$ and $\alpha$-YbAl$_{1-x}$Fe$_x$B$_4$ $(x=0.014)$. 

Figure~\ref{fig:nfnd}(b) shows the $\varepsilon_f$ dependences of the ED and MT moments at the Yb1 site, which are defined as $\bar{X}_{\alpha\zeta}\equiv \bar{X}_{\alpha\zeta+}+ \bar{X}_{\alpha\zeta-}$ with $\bar{X}_{\alpha\zeta\pm}=\frac{1}{N}\sum_{i}\langle X_{i\alpha\zeta\pm}\rangle$ for $X=Q, T$ and $\zeta=x, y$. 
The results of $\bar{Q}_{1x}=0$ and $\bar{Q}_{1y}
\ne
 0$ indicate that the 
ED
 moment along $y$ direction exists. 
Interestingly, $\bar{Q}_{1y}$ becomes zero in the vicinity of the QCP and changes sign, i.e., $\bar{Q}_{1y}>0$ $(\bar{Q}_{1y}<0)$ for $\varepsilon_{f}\lsim\varepsilon_f^{QCP}$ $(\varepsilon_{f}\gsim\varepsilon_f^{QCP})$, whose absolute value has a maximum at $\varepsilon_f=-2.294>\varepsilon_f^{QCP}$. The MT moments along $x$ direction for each Kramers state $\bar{T}_{1x\pm}$ also show the sign changes in the vicinity at the QCP, which are also enhanced for $\varepsilon_{f}>\varepsilon_{f}^{QCP}$ with opposite signs while $\bar{T}_{1y\pm}=0$. We note that the relation $|\bar{Q}_{1y}|=\frac{4e}{3\mu_{B}}|\bar{T}_{1x\pm}|$ holds.
Thus the total MT moment is zero $\bar{T}_{1\zeta}=0$ for $\zeta=x, y$. 
As for the Yb2 site, we obtained $\bar{Q}_{2x}=0$, $\bar{Q}_{2y}=-\bar{Q}_{1y}$, $\bar{T}_{2x}=0$, and $\bar{T}_{2y}=0$. These results are consistent with the perturbation analysis, which indicates that the odd-parity CEF [see Eq.~(\ref{eq:CEF_odd})] is actually generated in Eq.~(\ref{eq:EPAM}).

\begin{figure}[t]
\includegraphics[width=8cm]{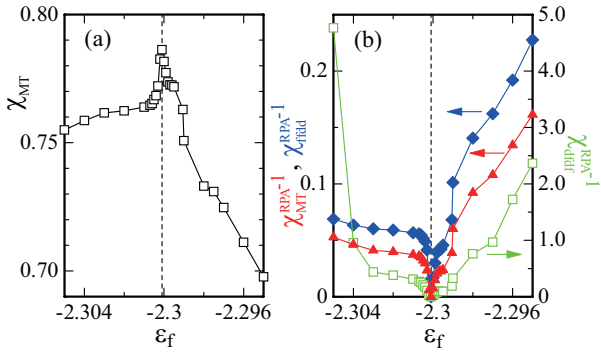}%
\caption{(color online) (a) $\chi_{MT}$ vs. $\varepsilon_{f}$ for $U_{fd}=U_{fd}^{QCP}$. 
(b) The $\varepsilon_f$ dependences of ${\chi_{MT}^{RPA}}^{-1}$ 
(filled triangle), 
${\chi_{ffdd}^{RPA}}^{-1}$ (filled diamond), and ${\chi_{dfdf}^{RPA}}^{-1}$ 
(square) for $U_{fd}=U_{fd}^{c}$ (see text).  
In (a) and (b), the dashed line indicates $\varepsilon_{f}=\varepsilon_{f}^{QCP}$ and we set $\mu_{B}=1$. 
}
\label{fig:MT_EfA}
\end{figure}

Although $\bar{\bm T}_{\alpha}={\bm 0}$ is the consequence of the time reversal symmetry of the paramagnetic (PM) state, the MT fluctuation can arise even in the PM state. 
Then, we calculate the susceptibility of the MT moment 
\begin{eqnarray}
\chi_{T_{x}T_{x}}({\bm q},\omega)
=\frac{i}{N}\int_{0}^{\infty}dt e^{i\omega t}
\langle[T^{x}_{\bm q}(t), T^{x}_{-\bm q}(0)]\rangle
\end{eqnarray}
with $T_{\bm q}^{x}=\sum_{i}e^{-i{\bm q}\cdot{{\bm r}_{i}}}T_{i1x}$. 
Figure~\ref{fig:MT_EfA}(a) shows the $\varepsilon_{f}$ dependence of $\chi_{MT}\equiv\lim_{{\bm q}\to{\bm 0}}\chi_{T_{x}T_{x}}({\bm q},0)$ for $U_{fd}=U_{fd}^{QCP}$. 
We find that $\chi_{MT}$ has a peak at $\varepsilon_f=\varepsilon_{f}^{QCP}$. 
This implies that the MT fluctuation is enhanced at the QCP. 
Since the CVF i.e. CT fluctuation diverges at the QCP, the MT fluctuation is expected to diverge at the QCP if the effect of the CVF is taken into account.  

To quantify this beyond the mean-field theory, let us rewrite the MT susceptibility in the form of  
%
\begin{eqnarray}
\chi_{T_{x}T_{x}}({\bm q},\omega)=
\frac{9\mu_{B}^2}{28}
\sum_{\nu=\pm}\left[\chi_{\nu\nu}^{ffdd}({\bm q},\omega)+\chi_{\nu\nu}^{ddff}({\bm q},\omega)
\right.
\nonumber
\\
\left.
-\chi_{\nu\nu}^{dfdf}({\bm q},\omega)-\chi_{\nu\nu}^{fdfd}({\bm q},\omega)\right],
\label{eq:chiTxTx}
\end{eqnarray}
%
where $\chi_{\nu\nu}^{\beta\gamma\delta\eta}({\bm q},\omega)$ is defined by 
$
\chi_{\nu\nu}^{\beta\gamma\delta\eta}({\bm q},\omega)\equiv\frac{i}{N}\int_{0}^{\infty}dt
e^{i\omega t}\langle[\Delta_{{\bm q}\nu}^{\gamma\delta}(t), \Delta_{{-\bm q}\nu}^{\eta\beta}(0)]\rangle
$
%
with 
$
\Delta_{{\bm q}\pm}^{fd}=\sum_{\bm k}
f^{\dagger}_{{\bm k}+{\bm q}1\pm\frac{5}{2}}d_{{\bm k}1\pm\frac{3}{2}}
$
and 
$
\Delta_{{\bm q}\pm}^{df}=\sum_{\bm k}
d^{\dagger}_{{\bm k}+{\bm q}1\pm\frac{3}{2}}f_{{\bm k}1\pm\frac{5}{2}}
$. 
Since we are currently considering the PM state without applying magnetic field so that $\chi^{\beta\gamma\delta\eta}_{++}=\chi^{\beta\gamma\delta\eta}_{--}$ holds, 
we omit the index $\nu$ in the expressions of the susceptibility hereafter. 
Near the QCP, the CT fluctuation caused by $U_{fd}$ is enhanced~\cite{Miyake2007}, which can be calculated by the 
random phase approximation (RPA)
 as the corrections for the mean-field state 
\begin{eqnarray}
\hat{\chi}({\bm q},\omega)=\hat{\chi}_{0}({\bm q},\omega)\{\hat{1}-\hat{U}\hat{\chi}_{0}({\bm q},\omega)\}^{-1},
\label{eq:RPA}
\end{eqnarray}
where $\hat{\chi}$, $\hat{\chi}_0$, and $\hat{U}$ are given by
\begin{equation}
\hat{\chi}_{(0)}({\bm q},\omega)=
\begin{bmatrix}
\chi^{ffdd}_{(0)}({\bm q},\omega) & \chi^{fdfd}_{(0)}({\bm q},\omega) \\
\chi^{dfdf}_{(0)}({\bm q},\omega) & \chi^{ddff}_{(0)}({\bm q},\omega)
\end{bmatrix},
\hat{U}=
\begin{bmatrix}
U_{fd} & 0 \\
0 & U_{fd}
\end{bmatrix},
\nonumber
\end{equation}
and $\hat{1}$ is the identity matrix. 
Here, the index 0 specifies the susceptibility calculated for the mean-field state. 
The critical point within this RPA formalism, defined by the point where ${\rm det}\{\hat{1}-\hat{U}\hat{\chi}_{0}({\bm q},\omega)\}=0$ is satisfied, is given by $U_{fd}^{c}=
1.635
$ and $\varepsilon_{f}=\varepsilon_{f}^{QCP}$. 
Then, by substituting $U_{fd}=U_{fd}^{c}$ and each $\varepsilon_{f}$ in the vicinity of $\varepsilon_{f}^{QCP}$ into Eq.~(\ref{eq:RPA}) and using the resultant $\chi^{ffdd}({\bm q},\omega)[=\chi^{ddff}({\bm q},\omega)]$ and $\chi^{dfdf}({\bm q},\omega)[=\chi^{fdfd}({\bm q},\omega)]$, we calculate the $\varepsilon_{f}$ dependence of $\chi_{MT}^{RPA}\equiv\lim_{{\bm q}\to{\bm 0}}\chi_{T_{x}T_{x}}({\bm q},0)$. 
As shown in Fig.~\ref{fig:MT_EfA}(b), the CT fluctuations by $\chi_{ffdd}^{RPA}\equiv\lim_{{\bm q}\to{\bm 0}}\chi^{ffdd}({\bm q},0)$ and $\chi_{dfdf}^{RPA}\equiv\lim_{{\bm q}\to{\bm 0}}\chi^{dfdf}({\bm q},0)$ diverge at $\varepsilon_{f}=\varepsilon_{f}^{QCP}$, which induce the divergence of the MT fluctuation, i.e., $\chi_{MT}^{RPA}=\infty$.

Since the MT moment is expressed in Eq.~(\ref{eq:MT}), it can have a finite matrix element between the 4f and 5d states with the magnetic quantum numbers $m$ and $m \pm 1$, as the ED moment. Hence, the $_{4f}\langle\pm\frac{5}{2}|$ and $|\pm\frac{3}{2}\rangle_{5d}$ states gave the finite values $\bar{T}_{\alpha x\pm}$ and $\bar{Q}_{\alpha y}$ at the locally inversion-symmetry broken Yb site. By the effect of the onsite 4f-5d Coulomb repulsion $U_{fd}$ at Yb, the CT fluctuation is enhanced, which eventually diverges at the valence QCP. This causes the divergences of the MT and ED fluctuations since they consist of the CT fluctuations as Eq.~(\ref{eq:chiTxTx}), whose operators are expressed as linear combinations of the CT-type operators as $f^{\dagger}_{i\alpha\pm\frac{5}{2}}d_{i\alpha\pm\frac{3}{2}}$ and its Hermitian conjugate. It is also noted that the Yb 5d states, which consist of the spherical harmonics $Y_{2,\pm 2}(\hat{\bm r})$ and/or $Y_{2,\pm 1}(\hat{\bm r})$, can have the finite matrix elements of the MT as well as the ED moment. Hence, the similar effects are expected to appear also for the other 5d states than the $|J=5/2, J_z=\pm 3/2\rangle$ state. 

Since the relation 
\begin{eqnarray}
\chi_{Q_{y}Q_{y}}({\bm q},\omega)=
\frac{4e^2}{9\mu_{B}^2}
\chi_{T_{x}T_{x}}({\bm q},\omega) 
\label{eq:Q_T}
\end{eqnarray}
 holds where $\chi_{Q_{y}Q_{y}}({\bm q},\omega)$ is defined as 
$
\chi_{Q_{y}Q_{y}}({\bm q},\omega)
=\frac{i}{N}\int_{0}^{\infty}dt e^{i\omega t}
\langle[Q^{y}_{\bm q}(t), Q^{y}_{-\bm q}(0)]\rangle
$ 
with $Q_{\bm q}^{y}=\sum_{i}e^{-i{\bm q}\cdot{{\bm r}_{i}}}Q_{i1y}$, the ED fluctuation along $y$ is proportional to the MT fluctuation along $x$. Hence, the ED fluctuation diverges simultaneously with the MT fluctuation at the valence QCP. 
Although the ED moments at the Yb1 and Yb2 sites [see Fig.~\ref{fig:Yb_B}(b)] are cancelled $\bar{Q}_{1y}=-\bar{Q}_{2y}$, their fluctuations exist in the PM state. Equation (\ref{eq:Q_T}) implies that the measurement of the ED fluctuations by the dielectric constant can detect the MT fluctuation. 
The detection of the MT as well as ED fluctuation by using various experimental probes such as X-ray diffuse scattering and M\"{o}ssbauer measurements is an interesting subject in the future. 

We also confirmed the tendency that even within the mean field theory, by applying conjugate field to the MT moment $-\sum_{i\alpha}h_{\alpha}T_{i\alpha x}$ to Eq.~(\ref{eq:EPAM}), $\partial \bar{T}_{\alpha x}/\partial h_{\alpha}$ diverges at the QCP for a finite $h_{\alpha}$. This also supports that the MT fluctuation diverges at the QCP. 

In $\alpha$-YbAl$_{1-x}$Fe$_x$B$_4$, the coincidence of the valence QCP and the magnetic transition seems to occur at $x=0.014$ as discussed in Ref.~\citen{WM2011}. This suggests a possibility that the MT order occurs for $x>0.014$ and the divergence of the uniform MT fluctuation with ${\bm q}={\bm 0}$ at the QCP affects the MT orders. In $\beta$-YbAlB$_4$, there is also a possibility that the MT order occurs. In that case, it is expected that the MT moment is aligned to $+x$~$(-x)$ direction at the Yb1 (Yb2) site, or vice versa, giving the alignment of the staggered MT moments in the $ab$ plane [see Fig.~\ref{fig:Yb_B}(b)]. 

In summary, by constructing the realistic minimal model for $\beta$-YbAlB$_4$, 
we have shown that the 4f-5d Coulomb repulsion under the odd parity CEF is the microscopic origin to induce the divergence of the MT fluctuation as well as the ED fluctuation simultaneously with the divergence of the CVF at the QCP of the valence transition. 
Our study has revealed the underlying mechanism by which novel multipole degrees of freedom can be active as fluctuations, which is a new aspect of the CT effect.

\begin{acknowledgment}
The authors thank H. Kusunose, S. Hayami, M. Yatsushiro, H. Kobayashi, and C. Bareille for useful discussions. 
This work was supported by JSPS KAKENHI Grant Numbers JP18K03542, JP18H04326, and JP17K05555.
\end{acknowledgment}

\end{document}